\documentclass[12pt,preprint]{aastex}

\def\kms{km~s$^{-1}$}

\def\kms{\mbox{km~s$^{-1}$}}

\def\cmq{cm$^{-2}$}

\def\um{$\mu$m}
\def\deg{$^{\circ}$}
\def\Msun{\mbox{$M_\odot$}}
\def\Lsun{\mbox{$L_\odot$}}
\def\Rsun{\mbox{$R_\odot$}}
\def\Vlsr{$V_{\rm LSR}$}
\def\Vsys{$V_{\rm sys}$}

\def\co{$^{12}$CO}

\def\Lbol{$L_{\rm bol}$}

\def\dVres{\mbox{$\Delta v_{\rm res}$}}

\def\V0{\mbox{$V_{\rm 0}$}}

\def\Vlsr{$V_{\rm LSR}$}
\def\Vsys{$V_{\rm sys}$}

\def\Vt{$V_{\rm t}$}

\def\Tr{\mbox{$T_{\rm rot}$}}
\def\Tr21{\mbox{$T_{\rm r,21}$}}
\def\Tmb{\mbox{$T_{\rm mb}$}}
\def\Tsys{\mbox{$T_{\rm sys}$}}
\def\Tmb{\mbox{$T_{\rm mb}$}}

\def\Tex{\mbox{$T_{\rm ex}$}}

\def\Teff{\mbox{$T_{\rm eff}$}}
\def\Fco{\mbox{$F_{\rm CO}$}}

\def\td{\mbox{$t_{\rm d}$}}

\def\td{\mbox{$t_{\rm d}$}}
\def\Fco{\mbox{$F_{\rm co}$}}

\def\Llobe{\mbox{$l_{\rm lobe}$}}
\def\Vt{\mbox{$V_{\rm t}$}}
\def\Vb{\mbox{$V_{\rm b}$}}
\def\Vflow{\mbox{$V_{\rm flow}$}}
\def\Mlobe{\mbox{$M_{\rm lobe}$}}

\def\thetahpbw{$\theta_{\rm HPBW}$}
\def\thetaeff{$\theta_{\rm eff}$}

\def\sgm{{$\sigma$}}

\def\lesssim{\mathrel{\hbox{\rlap{\hbox{\lower4pt\hbox{$\sim$}}}\hbox{$<$}}}}
\def\gtrsim{\mathrel{\hbox{\rlap{\hbox{\lower4pt\hbox{$\sim$}}}\hbox{$>$}}}}

\slugcomment{2009 July 28}

\shorttitle{High Velocity Outflow in CO $J=7-6$ from the Orion Hot Core}
\shortauthors{Furuya and Shinnaga}

\begin{document}


\title{High Velocity Molecular Outflow in CO $J=7-6$ Emission
from the Orion Hot Core}


\author{Ray S. FURUYA\altaffilmark{1}}
\affil{Subaru Telescope, National Astronomical Observatory of Japan}
\email{rsf@subaru.naoj.org}

\and

\author{Hiroko SHINNAGA\altaffilmark{2}}
\affil{Caltech Submillimeter Observatory, California Institute of Technology}
\email{shinnaga@submm.caltech.edu}


\altaffiltext{1}{650 North A'ohoku Place, Hilo, HI\,96720}
\altaffiltext{2}{111 Nowelo Street, Hilo, HI\,96720}


\begin{abstract}
Using the Caltech Submillimeter Observatory 10.4-meter telescope,
we performed sensitive mapping observations of $^{12}$CO $J=7-6$ 
emission at 807\,GHz towards Orion IRc2.
The image has an angular resolution of 10\arcsec,
which is the highest angular resolution data
toward the Orion Hot Core published for this transition.
In addition,
thanks to the on-the-fly mapping technique,
the fidelity of the new image is rather high,
particularly in comparison to previous images.
We have succeeded in mapping the northwest-southeast
high-velocity molecular outflow, 
whose terminal velocity is
shifted by $\sim 70-85$ \kms\
with respect to the systemic velocity of the cloud.
This yields an extremely short dynamical time scale of $\sim$900 years.
The estimated outflow mass loss rate shows an extraordinarily high value,
on the order of $10^{-3}$ \Msun\ yr$^{-1}$.
Assuming that the outflow is driven by Orion IRc2,
our result agrees with the picture so far obtained 
for a 20 \Msun\ (proto)star in the process of formation.
\end{abstract}





\section{Introduction}

Despite its astrophysical importance, the high-mass
($M_{\ast}\gtrsim 8$ \Msun ) star formation process remains
poorly understood both observationally and theoretically.
High-velocity (HV) molecular outflows are the most prominent 
phenomena seen, not only in high-mass but also in low-mass
star-forming regions.
Molecular outflows from low-mass stars are thought to be 
momentum-driven by highly collimated stellar winds,
and are closely related to the accretion process onto
the central star (e.g., Arce et al. 2007, and references therein).
Although such a paradigm cannot simply be scaled up to the
high-mass regime, 
one may obtain important information about the formation mechanism
of early-type stars by observing their outflows.
In this context, the Orion Hot Core (HC) is
the best site to study massive star formation process, 
as its proximity allows very detailed studies.\par

Ground-based mapping observations of CO lines at submillimeter
(submm) wavelengths 
towards the region were pioneered by, e.g., 
\citet{w01} in the $J=7-6$ transition and
\citet{marrone04} in $J=9-8$.
These authors focused on studying the properties of
the intense low-velocity gas around the velocity of 
the natal molecular cloud,
which arises predominantly from a photo-dissociation region (PDR).
Observations of low-$J$ lines, such as $J=2-1$,
allow observers to attain high angular resolution images 
down to $\sim$1\arcsec\
by utilizing interferometers, as demonstrated by
\citet{chernin96}, \citet{beuther08}, and others.
However, such low-$J$ line observations of molecular outflows
usually make it difficult to distinguish low-velocity outflowing gas
from the quiescent ambient gas.
The $J=7$ level is approximately 156\,K above the ground-state level,
and thus more sensitive to the presence of hot ($T>100$\,K) gas.
The Einstein A coefficient for the $J=7-6$ transition is 50 times
greater than for the $J=2-1$ line,
providing a high contrast between the warm and cold gas. 
Such characteristics greatly help us to distinguish HV outflowing gas
from the surrounding material.
To characterize the physical properties of the 
outflowing gas in the Orion HC, 
we performed observations of CO $J=7-6$ using 
the Caltech Submillimeter Observatory 
(CSO)\footnote{Caltech Submillimeter Observatory is 
operated by the California Institute of Technology under the grant from
the US National Science Foundation (AST\,08-38261).} 
10.4-meter telescope, 
making full use of its high spatial and high
spectral resolution.

\section{Observations and Data Reduction}
\label{s:obs}

We carried out on-the-fly (OTF) mapping observations
of the CO $J=$7--6 line 
(rest frequency, $\nu_{\rm rest}$ =\,806651.720\,MHz) 
with the CSO on 2009 January 21 and 22.
The observations took 1--2 hrs each day.
Prior to the submm observations, we performed optical pointing
observations on 2009 January 20.
Therefore, we believe that the overall pointing accuracy 
was better than 3\arcsec .
We used the 850\,GHz receiver \citep{kooi00} with
the Fast Fourier Transform Spectrometer (FFTS) as a back-end, 
yielding an effective velocity resolution
(\dVres ) of 0.0454 \kms\ with the 1\,GHz bandwidth mode.
During the observations, the atmospheric zenith optical depth 
at 225\,GHz ranged between 0.035 and 0.055, 
and the single side-band (SSB) system temperature (\Tsys )
stayed between $\sim$2500\,K and 8000\,K,
depending on the optical depth and the airmass.
We wish to point out that our observations are almost twice as
sensitive as the previous \co\ (7--6) observations by \citet{w01}, 
whose typical \Tsys\ was $\sim 14,000$\,K.
The OTF mapping was carried out over an area of $\sim 170\arcsec$
centered on Orion IRc2 
(R.A. $=5^h35^m14.5^s$, Dec. $=-$5\deg\ 22\arcmin\ 30\farcs4 in B1950)
by scanning once each along the R.A. and Dec. directions.
Subsequently, we decided to concentrate on mapping 
the central 85\arcsec\ region,
which we further scanned twice along the R.A. and once along 
the Dec. directions.
The OTF mapping scans were gridded with a pixel size scale of 5\farcs0 by
employing a position-switching method.
The telescope pointing was checked by observing the continuum emission
towards Venus and Saturn every hour.
At the line frequency, the beam size (\thetahpbw ) is estimated to be
10\arcsec\, and the main-beam efficiency ($\eta_{\rm mb}$) was 0.34
from our measurements towards Saturn.
All the spectra were calibrated by the standard chopper wheel method, 
and were converted to main-beam brightness temperature
(\Tmb ) scale by dividing by $\eta_{\rm mb}$.
The uncertainty in the intensity calibration is estimated to be $\sim$\,20\%. 
After calibrating all the spectra,
we made two 3D data cubes from two groups of the spectra taken by 
scanning along R.A. and Dec. directions.   
These two cubes were processed with the code ``Basket-Weave'', 
implemented in the NOSTAR package (Sawada et al. 2007) to remove
the scanning effect using the method developed by 
Emerson \& Graeve (1988).

\section{Results and Discussion}
\label{s:results}

\begin{figure}
\begin{center}
\includegraphics[angle=-90,scale=.65]{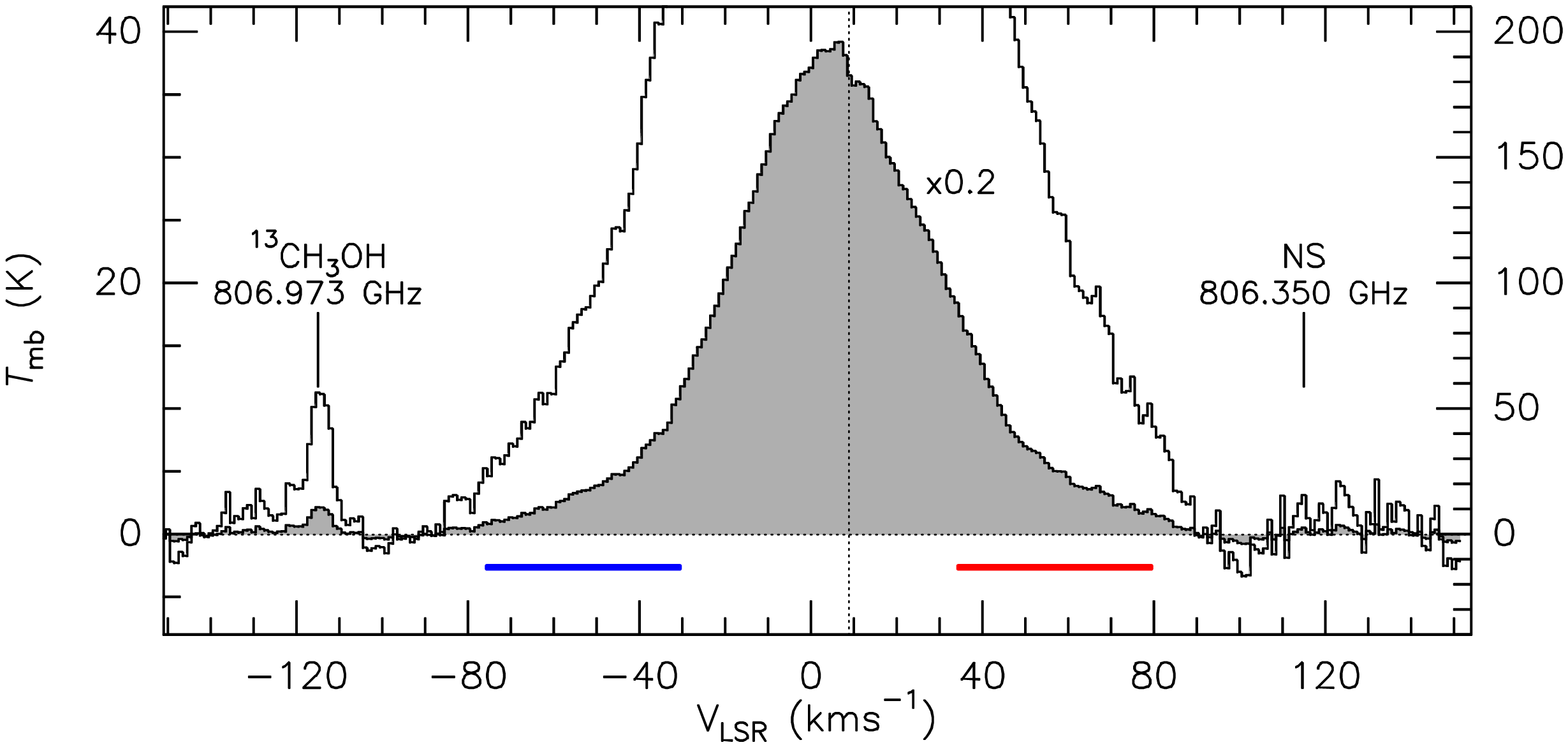} 
\caption{Single-dish spectra of the $^{12}$CO $7-6$ emission toward
Orion IRc2 in the main-beam brightness temperature (\Tmb) scale
with a velocity resolution of 0.726 \kms.
See the $y$-axis labels of the right- and left-hand sides
for the overall (filled histogram) and the magnified spectra, respectively.
The RMS noise level is 0.72\,K for 81 seconds integration.
The blue- and red-colored horizontal bars below the spectra 
indicate the LSR-velocity ranges used to obtain the integrated 
intensity map shown in Figure \ref{fig:hvmaps}.
The systemic velocity of the cloud is \Vlsr\ $\sim\,9$ \kms\ 
which is indicated by the vertical dashed line.
\label{fig:sp}}
\end{center}
\end{figure}

Figure \ref{fig:sp} shows our CO (7--6) spectrum towards
Orion IRc2 in the \Tmb\ scale.
The spectral profile is rather broad, 
and shows prominent HV wing emission. 
The HV wing is seen both blue- and redshifted sides up to 
$\sim$80 \kms\ away from the systemic velocity
(\Vsys ) of the cloud, \Vlsr $= 9$ \kms.
The peak \Tmb\ is 202\,K at \Vlsr $=\,1.9$ \kms, which 
is consistent with that measured by \citet{c05},
although the pointing centers differ between our maps 
and theirs by 5\farcs0.
The CO spectrum shows a dip at \Vsys, 
suggesting that the line is optically thick around \Vsys.
The two isolated emission lines can be seen at 
\Vlsr $\sim +110$\,\kms\ and $\sim -115$ \kms; 
they are identified
as lines of NS ($f_{\rm rest} =$\,806350\,MHz)
and $^{13}$CH$_3$OH (806973\,MHz), on the basis of the 
line survey \citep{c05} at the 350 \micron\ band.\par

\begin{figure}
\begin{center}
\includegraphics[angle=-90,scale=.85]{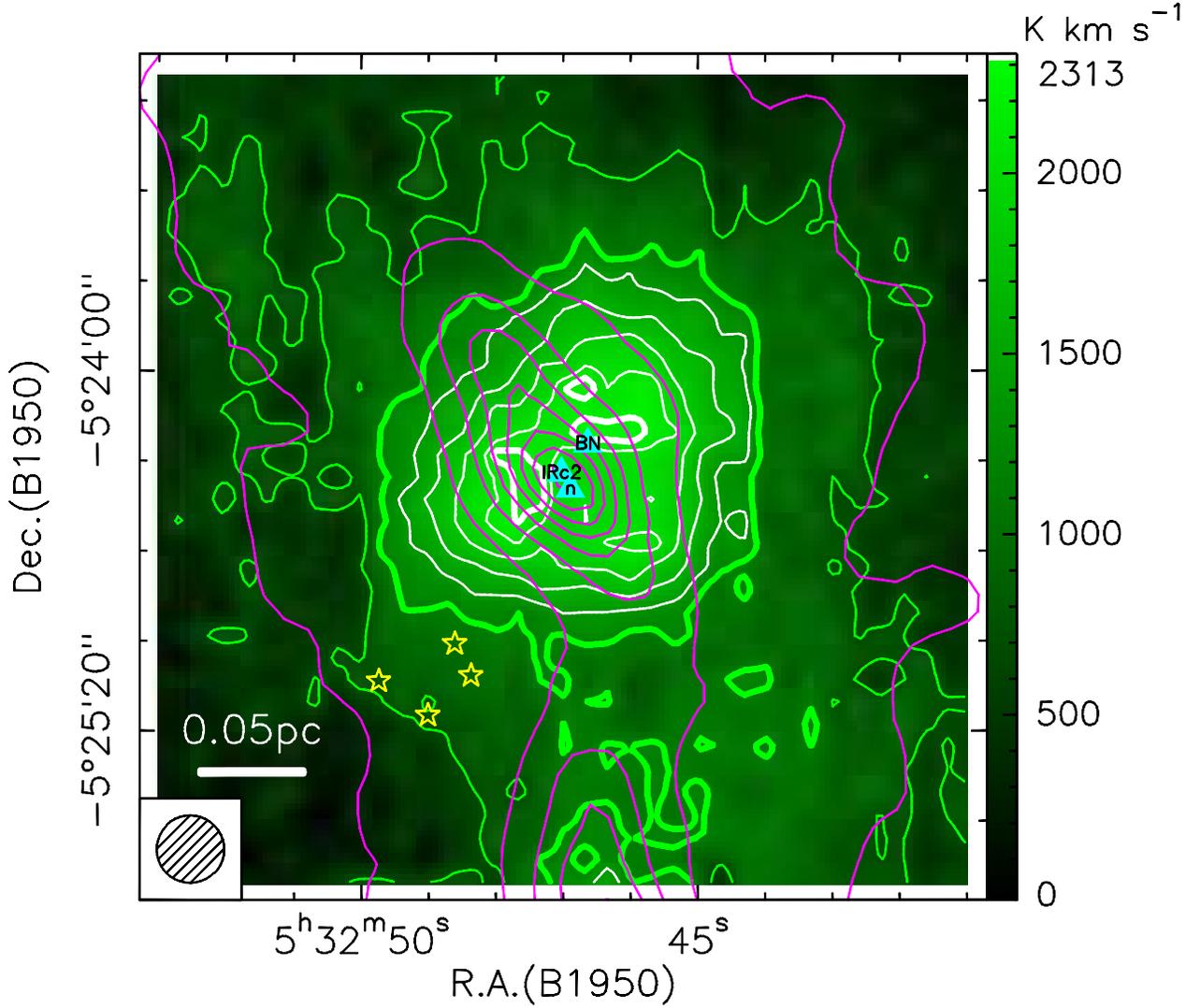} 
\caption{Overlay of the bulk emission of the \co\ (7--6)  
(green image plus white and green contours) on the
350 \um\ continuum emission (magenta contours; Lis et al. 1998).
The contours for the \co\ (7--6) emission are the 95\% level
(thick white), 90\%, 80\%, 70\%, 60\% (thin white),
50\% (thick green), and 30\% (thin green) level of the peak
for the clarity of the plot.
The CO map is obtained by 
integrating the emission between \Vlsr $=$ 4.5 \kms\ and 14.5 \kms.
The light blue triangles with source names show the positions of the radio
continuum sources [\citet{mr95}; \citet{rzh09}], 
and the yellow stars indicate the
positions of the Trapezium members.
Note that the peak position of the infrared source IRc2 is displaced from
the radio source I by 0\farcs8 \citep{gezari92}.
The hatched ellipse at the bottom left corner indicates
the HPBW of the CSO beam (10\arcsec ).
\label{fig:maps}}
\end{center}
\end{figure}

Figure \ref{fig:maps} shows an overlay of the low-velocity
bulk emission of the CO (7--6) and the 350 \micron\ continuum 
emission \citep{lis98}. 
The (7--6) line shows intense emission surrounding IRc2, 
and spreads along the north-south direction if we consider
the weak extended component.
The CSO beam is too large to resolve the spatial
separation of $\sim$0\farcs8 between IRc2 and Source I \citep{gezari92} 
which has been proposed as a powering
source for the Orion outflow 
(Beuther \& Nissen 2008, and references therein).
The north-south elongation of the weak emission 
corresponds to the Orion Ridge;
the northern tip of the Orion South core can be recognized 
at the bottom (i.e., south) of our map.
The widespread quiescent CO (7--6) emission is thought to 
be associated with the PDR heated by $\Theta ^1C$ Orionis \citep{w01}.
The bulk emission shows a condensation with a diameter of 
$\sim 80\arcsec$, measured at the 50\% level contour, 
corresponding to 0.065 pc at $d=450$\,pc.
It should be noted that the CO (7--6) line and the 350 \um\ continuum
emission do not show similar spatial structure; the former has a roundish
shape, whereas the latter is elongated along the north-south direction.
Another result deduced from Figure \ref{fig:maps} is that 
the bulk emission does not show a single peak at the position 
of the 350 \um\ continuum peak.
Instead, it shows two local maxima; one of them lies to the west of
the BN object and the other is $\sim 10\arcsec$ east of IRc2.
Positions of the two local maxima
are roughly consistent with the peak positions
of the blue- and redshifted CO (7--6) wing emission reported in 
\citet{w01}, 
although their HV wing emission maps (see their 
Figure 5) show a scanning effect.\par

\begin{figure}
\includegraphics[angle=-90,scale=.72]{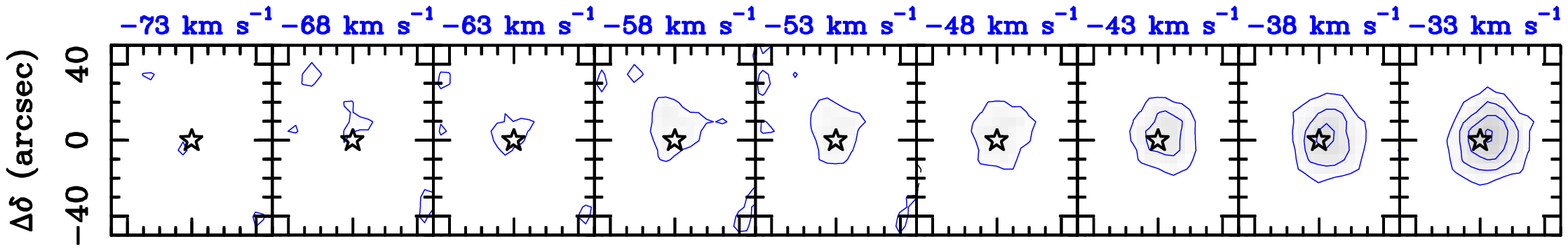} 
\includegraphics[angle=-90,scale=.72]{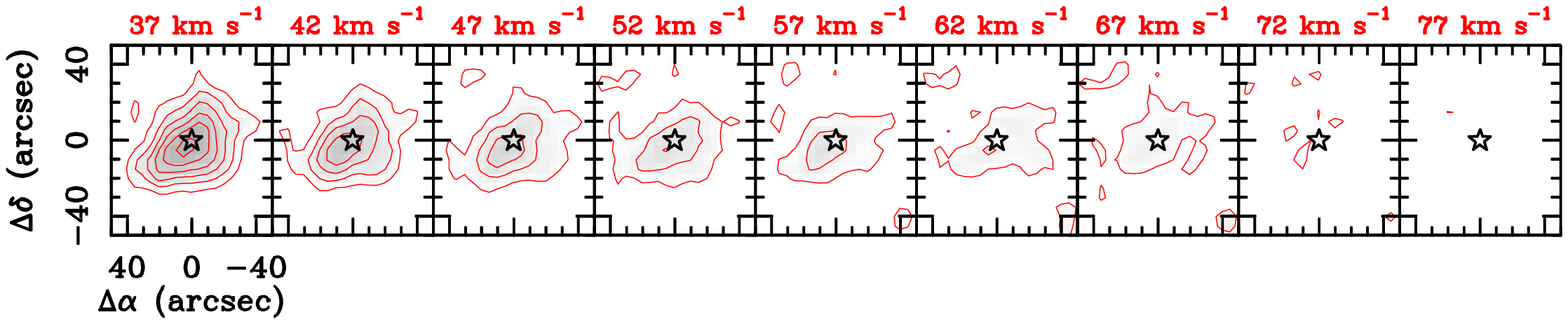} 
\caption{Velocity channel maps of the high-velocity \co\ (7--6) emission toward the
Orion IRc2 (the star) with an effective angular resolution of 13\arcsec\ (see text).
Each channel map is averaged over a 5.0 \kms\ bin whose central 
LSR-velocity is shown on the top of each panel.
All the contours are 2\sgm\ intervals, increasing from the 3\sgm\ level
where 1\sgm\ is 11\,K in \Tmb.
\label{fig:chmaps}}
\end{figure}

Figure \ref{fig:chmaps} shows velocity channel maps of 
the HV emission. 
To produce the channel maps, we convolved the data with 
a Gaussian beam so that the effective beam size (\thetaeff )
becomes 13\arcsec, to obtain a higher signal-to-noise ratio (S/N).
Here, the choice of 13\arcsec\ is to compare with the
previous work by \citet{w01}, who used the HHT 10-meter telescope.
The channel maps show that 
the redshifted gas is elongated to the (south)east of the central star,
whereas the blueshifted gas seems to be
distributed westward.\par

\begin{figure}
\begin{center}
\includegraphics[angle=-90,scale=.60]{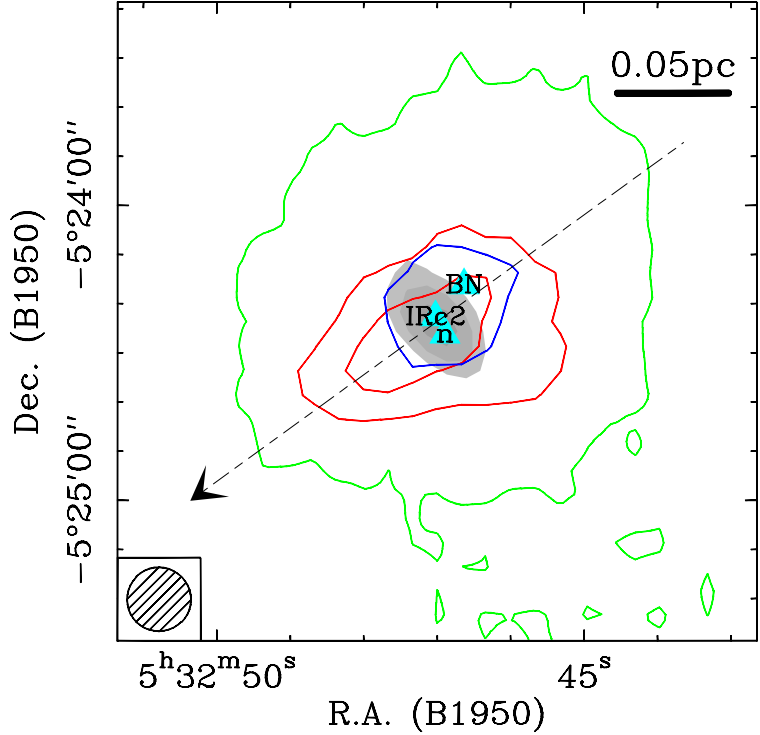} 
\caption{Comparison of the blue- and redshifted $^{12}$CO (7--6) 
high-velocity (HV) wing emission.
The green contour, taken from Figure \ref{fig:maps}, 
represents the 50\,\% level contour of the CO bulk emission, 
and the greyscale shows the 350 \um\ continuum emission \citep{lis98}.
The contours for the HV emission are 2$\sigma$ intervals, 
increasing from the $4\sigma$ level 
where $\sigma =$ 10.6 and 9.9 K$\cdot$ kms\ for the blue- and redshifted
wings, respectively.
The blue- and redshifted emission are integrated over the velocity ranges of
$-75.5 \leq$\Vlsr/\kms $\leq -30.5$, and
$+34.5 \leq$\Vlsr/\kms $\leq +79.5$, respectively
(see the horizontal color-coded bars in Figure \ref{fig:sp}).
The hatched circle at the bottom-right corner indicates the
effective beam size (\thetaeff ) of 13\arcsec. 
The dashed line indicates the cut axis for the 
position-velocity diagram shown in Figure \ref{fig:pv}.
The other symbols are the same as those in Figure \ref{fig:maps}.
Notice that the possible heating sources discussed by
\citet{deVicente02} are located within a region of $\sim$1\farcs5 radius,
whose spatial-scale is almost comparable with 
the blue-triangle representing the IRc2 position.
\label{fig:hvmaps}}
\end{center}
\end{figure}

\begin{figure}
\begin{center}
\includegraphics[angle=-90,scale=.48]{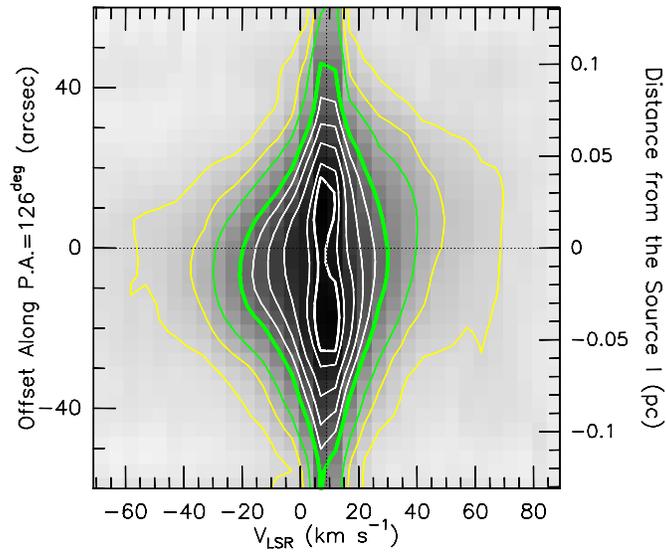}
\caption{Position-velocity (PV) diagram of the 
$^{12}$CO (7--6) emission along
the outflow axis (P.A.$=+126$\deg) passing through the position of IRc2
(see Figure \ref{fig:hvmaps}).
The white and green contours are the same as those in Figure \ref{fig:maps}, 
and the two yellow contours indicate the 
20\% and 10\% levels with respect to the peak intensity.
The vertical dashed-line at \Vlsr\ $=$ 9 \kms\ shows the \Vsys.
\label{fig:pv}}
\end{center}
\end{figure}

The distribution of the blue- and redshifted HV wing emission is 
compared in Figure \ref{fig:hvmaps} by overlaying the integrated
intensity maps.
Clearly, the HV gas is confined to the innermost region 
of the core, 
inside the 50\% level contour of the low-velocity bulk emission.
Here, we integrated the HV emission over the velocity ranges shown in 
Figure \ref{fig:sp}, i.e., 
over the velocity range between the boundary velocity (\Vb ),
which divides the LSR-velocity range of the outflowing gas from
the bulk emission, 
and the first highest LSR-velocity 
where the wing drops below the 5\sgm\ level.
We refer to such an LSR-velocity as the terminal velocity (\Vt ).
Since it is almost impossible to separate low-velocity outflowing
gas from the bulk ambient gas, 
we arbitrarily estimated \Vb\ from Figures \ref{fig:sp} and \ref{fig:chmaps}.
The redshifted gas shows an extended structure along the east-west direction,
whereas the blueshifted emission arises from a compact condensation, 
rather than an elongated structure.
It appears that the distribution of the blue- and redshifted 
gas is symmetric with respect to the peak position of the
350 \um\ continuum source, 
although the blue- and redshifted gas overlap each other.
We prefer to interpret this as representing the pair of 
molecular outflow lobes previously reported 
[\citet{chernin96}; \citet{rf99}; \citet{deVicente02}; \citet{beuther08}].
This is because their terminal velocities are too high to be
interpreted in the context of the other scenarios,
such as rotation or/and infall.
Given its position, we exclude the BN object as a candidate 
source for the outflow.
However, due to the limited angular resolution, 
our data do not give further constraints on identifying 
the driving source, as discussed by e.g., 
\citet{deVicente02}, \citet{beuther08}, \citet{rzh09},
and others.
Assuming that the position of IRc2 represents
the center of the outflow, we made a position-velocity (PV) diagram
of the CO (7--6) emission along the outflow axis (Figure \ref{fig:pv}).
The PV diagram principally confirms our findings described above.
Unfortunately, the CSO beam is too large to provide more detailed information
about, e.g., driving mechanisms for outflow and/or outflow history.\par

Subsequently, we estimated the outflow lobe mass (\Mlobe ) to
calculate kinematical properties 
such as mass-loss rate ($\dot{M}_{\rm flow}$) 
and momentum rate (\Fco).
Here, we assumed that the HV gas is optically thin and in LTE
because neither isotope line, i.e., $^{13}$CO $J=7-6$,
nor similar high-$J$ lines, e.g., $^{12}$CO $J=6-5$, 
are available with the 10\arcsec\ beam.
We obtained \Mlobe\ $=$\,$(1.3\pm 0.3)$ and $(3.5\pm 0.9)$ \Msun\ 
for the blue and red lobes, respectively, 
by obtaining a mean column density over the lobes of 
$(9.0\pm 2.4)\times 10^{17}$ and $(1.0\pm 0.3)\times 10^{18}$\,\cmq.
Here, we adopted an excitation temperature (\Tex ) of 150\,K \citep{c05}, 
and used a $^{12}$CO/H$_2$ abundance ratio of $10^{-4}$ \citep{d78}.
For comparison with previous publications, we adopted $d=\,450$ pc,
although recent astrometry using the VLBI technique
has revised the distance to the region 
(414$\pm$7 pc, Menten et al. 2007; 
437$\pm$19 pc, Hirota et al. 2007).
It is likely that the outflow masses are underestimated because 
of the uncertainty in defining \Vb.
Since the highest outflow velocity is given by 
$\Vflow=|V_{\rm t}-V_{\rm sys}|$,
the dynamical time scale (\td) is estimated from \Llobe/\Vflow,
where the lobe length \Llobe\ is defined as the maximum extent of
the lobe measured from the IRc2 position.
We assumed that the outflow inclination angle ($i$) is 45\deg, 
here defined as the angle between the outflow axis and the line of sight.
The estimated \td\ $\sim$ 900\,years suggests that 
the outflow is extremely young.
Hereafter, we describe the mean values of the outflow properties,
because the derived outflow properties for the blue- and redshifted lobes
are comparable within the errors.
We can estimate the outflow mass-loss rate from
$M_{\rm lobe}$/\td.
This is $\sim 4\times 10^{-3}$ \Msun\ yr$^{-1}$ for both lobes.
Since molecular outflows appear to be momentum-driven \citep{cb92}, 
the momentum rate \Fco=\Mlobe$V_{\rm flow}^2$/\Llobe\
may be taken as an indicator of the outflow strength
and hence of the mass and luminosity of the young stellar 
object (YSO) powering it.
The outflow lobe has $\Fco\simeq 0.3$ \Msun\ \kms\ year$^{-1}$.
The derived properties are not affected significantly 
if one takes into account the unknown inclination angle $i$,
which is defined as the angle between the outflow axis
and line of sight.
In fact, \td, $\dot{M}_{\rm flow}$, and \Fco\
are corrected with factors of
$\cot i$, $\tan i$, and $\sin i/\cos ^2i$, respectively
(see e.g., Davis et al. 1997).
Thus, even if we assume extreme values of $i\sim20\degr$ 
and $70\degr$,
the corrections for the three quantities above would only be a factor 
of 3--5, thus leaving our estimates unaffected.
All the derived outflow parameters are extraordinarily large, 
implying that the powering YSO has 
a bolometric luminosity of $\sim 10^5-10^6$ \Lsun, 
if we apply the empirical relationship between the two
quantities (see Figure~5 of Richer et al. 2000 and 
Figure~4 of Beuther et al. 2002).\par

Our estimate of $\dot{M}_{\rm flow}$ to be on the order of
$10^{-3}$ \Msun\ yr$^{-1}$ 
strongly suggests that the central massive YSO is
gaining mass toward its final stellar mass through 
a disk-outflow system.
We do not exclude the possibility that mass accretion
onto the central star (or/and disk system) is currently enhanced.
A statistical study of molecular outflows in 26 high-mass star-forming
regions by \citet{beuther02} suggests that the ratio between 
the jet mass-loss rate and mass-accretion rate 
($\dot{M}_{\rm acc}$) may be $\sim 0.2$.
If this is the case, the accretion rate onto the outflow powering source
would be on the order of $\sim 10^{-2}$\Msun\ yr$^{-1}$.
Although the driving source of the Orion outflow has not been
firmly identified, and has been a matter of debate
(de Vicente et al. 2002; Beuther and Nissen 2008; Rodr\'iguez et al. 2009), 
it is tempting to suggest that
this accretion rate does not contradict 
that estimated by \citet{nakano00},
who proposed that the central massive YSO in Orion IRc2 is 
either a protostar with a stellar radius of $R_{\ast}\gtrsim 300$\,\Rsun\ 
or a protostar which has a disk with 
$\dot{M}_{\rm acc}\approx 10^{-2}$ \Msun\ yr$^{-1}$.
\citet{nakano00} used a one-zone model to account for the observational results
of $R_{\ast}\gtrsim 300$\,\Rsun\ and an effective temperature of 
\Teff $\sim$ 3000 -- 5500\,K \citep{morino98};
they adopted the latter interpretation because their
model could not reproduce an \Lbol\ $\gtrsim 4\times 10^4$ \Lsun\ (proto)star
(Kaufman et al. 1998; Gezari et al. 1998)
using the low \Teff\ and a stellar radius larger than 30 \Rsun.
More stringent calculations recently performed by \citet{ho09}
considered the possibility that such a massive protostar can be formed
if $\dot{M}_{\rm acc}$ is higher than 
$4\times 10^{-3}$ \Msun\ yr$^{-1}$.
If this is the case, the Keplerian speed at the protostellar surface
($v_{\rm kep} = \sqrt{GM_{\ast}/R_{\ast}}$), which is believed to 
represent the jet velocity (e.g., Richer et al. 2000; Arce et al. 2007),
becomes $\sim 220$ \kms\ for a 25 \Msun\ and 100 \Rsun\ star.
Here, the stellar mass and radius are taken from \citet{ho09}.
We believe that our result of \Vflow\ $\sim$ 80 -- 90 \kms\
agrees with the picture of Orion IRc2 so far obtained,
especially when we consider that 
molecular outflows are made of ambient material
entrained by a jet/winds from the central star.
In conclusion, our sensitive and high-fidelity \co\ (7--6) line imaging
demonstrates that a massive YSO is driving an extraordinarily powerful
molecular outflow, suggestive of a large accretion rate.
Clearly, higher resolution and better sensitivity imaging 
of the submm transition, as well as identification of its
driving source, are required to study the spatial structure of
the outflow. It may retain a history of the outflow events,
a clue to understanding the mass accretion history.





\acknowledgments

The authors gratefully acknowledge
Richard A. Chamberlin, Brian Force, and Hiroshige Yoshida
for their generous help during the observations, 
Thomas G. Phillips for his continuous encouragement, and
Darek C. Lis for providing the 350 \um\ continuum image.
The authors sincerely thank Thomas A. Bell
for his critical readings of the manuscript at the final stage of 
preparation.
R. S. F. wishes to thank Takashi Hosokawa for fruitful discussion.
This work is partially supported by a Grant-in-Aid
from the Ministry of Education, Culture, Sports, 
Science and Technology of Japan
(No.\,20740113).



{\it Facilities:} \facility{Caltech Submillimeter Observatory 10.4-meter telescope}

\clearpage



\end{document}